\documentclass[aps,floats,twocolumn,showpacs]{revtex4-1}
\usepackage[utf8]{inputenc}
\usepackage[english]{babel}
\usepackage{amssymb}
\usepackage{amsmath}
\usepackage[pdftex]{graphicx}
\usepackage{cmap}
\usepackage[pdftex]{hyperref}
 \def\bc{\begin{center}}          \def\ec{\end{center}}
 \allowdisplaybreaks

\begin{document}
 \title{ Generation of high-field narrowband terahertz radiation by counterpropagating plasma wakes
}
 \author{I.V.Timofeev, V.V.Annenkov, E.P.Volchok}
 \affiliation{Budker Institute of Nuclear Physics SB RAS, 630090, Novosibirsk, Russia \\
 Novosibirsk State University, 630090, Novosibirsk, Russia}
 \begin{abstract}

It is found that nonlinear interaction of plasma wakefields driven by counterpropagating laser or particle beams can efficiently generate high-power electromagnetic radiation at the second harmonic of the plasma frequency. Using a simple analytical theory and particle-in-cell simulations, we show that this phenomenon can be attractive for producing high-field ($\sim 10$ MV/cm) narrowband terahertz pulses with the gigawatt power level and millijoule energy content.

 \end{abstract}
 \maketitle

The terahertz radiation ($0.1-30$ THz) is now of great importance due to its numerous applications in science and technology. High-power sources in this frequency range open up novel opportunities in controlling molecular rotations, lattice vibrations and spin waves in matter \cite{Kampfrath2013}. Tremendous progress has been recently demonstrated in generating high-field ($10-100$ MV/cm) single-cycle terahertz pulses by using very different schemes such as difference frequency generation with two distinct lasers \cite{Sell2008}, optical rectification of single femtosecond laser pulses in nonlinear crystals \cite{Huang2013,Vicario2014}, transition radiation of high-energy electron beams \cite{Leemans2003,Wu2013}, irradiation of solid targets by relativistic intense lasers \cite{Gopal2013,Liao2016}, two-color laser interaction with gaseous \cite{Oh2014,Meng2016} and clustered \cite{Jahangiri2013} plasmas. 

Narrowband terahertz sources are also of great interest with regard to the resonant control and manipulation of matter, but generation of multi-cycle terahertz pulses with even moderate fields (1 MV/cm) and mJ energies still remains a challenging problem. Today, the most intense ($\sim 1$ MW, tens of $\mu$J) pulses of this radiation are produced by large-scale accelerator facilities such as free electron lasers \cite{Vinokurov2011,Shen2011,Li2016}. Table-top generation schemes providing $\mu$J pulses with narrow linewidths ($2-3\%$) are based on optical rectification of a temporally modulated chirped pump laser in an organic crystal \cite{Vicario2016} or in a periodically poled lithium niobate \cite{Carbajo2015}. Plasma is also considered as a promising nonlinear medium for generating  high-power multi-cycle terahertz pulses. It supports long-lived oscillations with extremely large electric fields and allows to tune the radiation frequency by a simple change of plasma density.

It has been recently proposed a number of generating schemes utilizing conversion of plasma oscillations to the terahertz electromagnetic (EM) waves. In particular,  plasma waves driven by laser or particle beams can produce radiation due to the linear mode conversion in a macroscopically inhomogeneous plasma \cite{Sheng2005} or via the antenna mechanism in a thin plasma with a small-scale longitudinal density modulation \cite{Timofeev2015,Annenkov2016a,Annenkov2016}. Plasma wakefields can also generate EM waves in external magnetic fields imposed along \cite{Wang2015} or across \cite{Yoshii1997,Yugami2002,Cho2015} the plasma column. In these schemes, however, plasma inhomogeneties widen the frequency spectrum of emitted radiation, and the magnetic field required for the upper frequency part of the terahertz range becomes too strong to be easily implemented in experiments. In this Letter we propose to generate high-field narrowband terahertz radiation by counterpropagating plasma wakes excited in a uniform plasma by short laser drivers. Such a scheme can generate GW, mJ multi-cycle terahertz pulses with the energy conversion efficiency higher than $10^{-4}$.

Let us first study the mechanism of EM radiation produced by counterpropagating plasma wakes independently on the driver nature. This mechanism is similar to that recently discussed in Ref. \cite{Annenkov2016}. Nonlinear interaction of two potential plasma waves oscillating with the plasma frequency and opposite longitudinal wavenumbers ( $(\omega_p,k_1)$  and $(\omega_p,-k_2)$) can generate the superluminal wave of electric current $(2\omega_p, k_1-k_2)$ which, in the bounded plasma, can pump vacuum EM waves. If these plasma wakefields are excited by relativistic drivers with the velocity $v_d\approx c$, the longitudinal wavenumber of the generating current is canceled and the generated EM waves escape transversely from the plasma channel. To calculate the radiation power from this finite-size plasma channel, we will follow the formalism described in Ref. \cite{Timofeev2016}. First, we will solve the problem in which amplitudes of plasma wakes are assumed  uniform along the channel and then will generalize the results to the case of a real laser filament with the slowly varying transverse structure. 

\begin{figure*}[htb]
\bc\includegraphics[width=500bp]{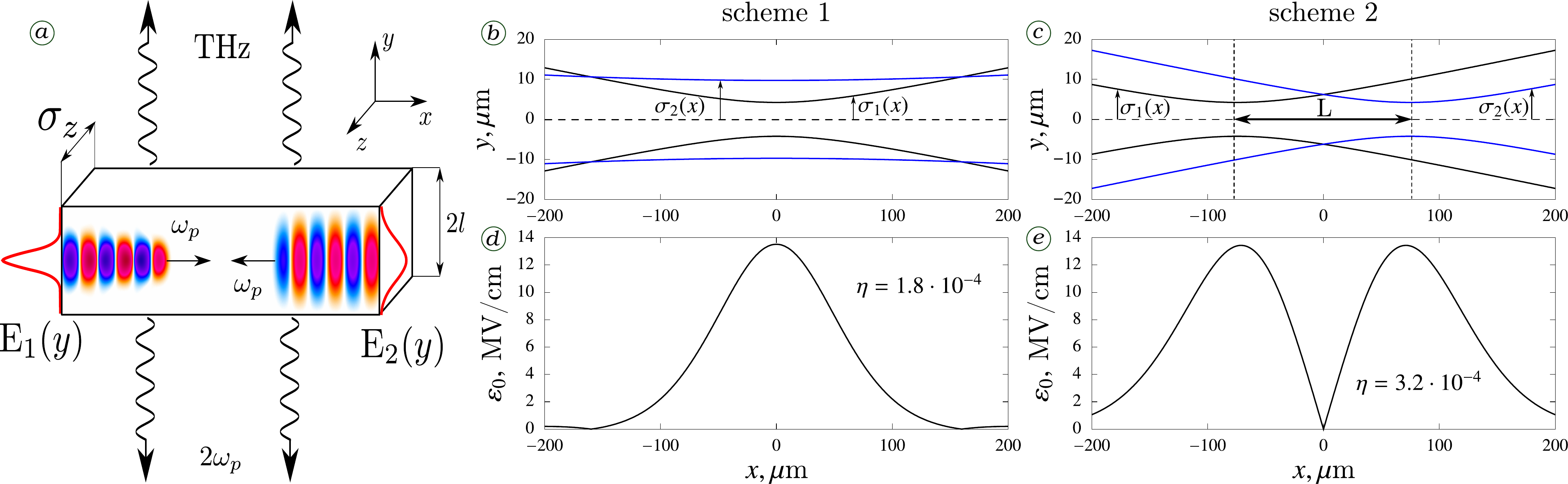} \ec \caption{Geometry of the problem (a). Two generating schemes with counterpropagating lasers: varying transverse sizes of laser beams (b, c); longitudinal profiles of the THz field $\mathcal{E}_0(x)$ (d,e) for the optimal parameters at the radiation frequency $\nu=31$ THz.}\label{f1}
\end{figure*}

In a plane plasma slab (Fig. \ref{f1}a), superposition of wakefields traveling in opposite directions along the coordinate $x$,
\begin{align}
	E_x&=\frac{1}{2}\left[\left(E_1(y)e^{i k x}+ E_2(y) e^{-i k x}\right) e^{-i \omega_p t}+ c.c.\right], \\
	E_y&=\frac{1}{2}\left[-\frac{i}{k}\left( E_1^{\prime} e^{ikx}- E_2^{\prime} e^{-ikx}\right) e^{-i\omega_p t} +c.c.\right],
\end{align}
generates  the nonlinear electric current,
\begin{equation}
	\delta j_x=j_0(y)e^{-i 2 \omega_p t} +c.c.,
\end{equation}
which oscillates with the doubled plasma frequency and does not depend on the longitudinal coordinate (here, the prime denotes the derivative with respect to $y$, $k=\omega_p/v_d$, $\omega_p=(4\pi e^2 n_0/m_e)^{1/2}$ is the plasma frequency, $n_0$ is the unperturbed plasma density, $e$ and $m_e$ are the charge and mass of an electron). In dimensionless units, when spatial sizes are measured in $c/\omega_p$, wavenumbers in $\omega_p/c$ and electric fields in $m_e c \omega_p/e$, the current amplitude can be written in the form
	\begin{equation}
		\mathcal{J}=\frac{j_0}{e n_0 c}=\frac{1}{4 } \left(E_1 E_2^{\prime\prime}-E_2 E_1^{\prime\prime}\right),
	\end{equation}
	where we neglect the difference between the driver velocity $v_d$ and the speed of light $c$ and put $k=1$.
	It is seen that this current density does not vanish and can produce transversely propagating EM radiation if the transverse structure of the first plasma wave differs from the similar structure of the second wave ($E_1(y)\neq E_2(y)$). Inside the plasma, the radiation field, $E_x=\mathcal{E}(y)e^{-i2 \omega_p t}+c.c.$, generated by this current is polarized along the channel and its amplitude should satisfy the equation
\begin{equation}\label{r}
	\mathcal{E}^{\prime\prime}+4\epsilon(2 \omega_p) \mathcal{E}=-2i\mathcal{J},
\end{equation}
 where $\epsilon(2\omega_p)=3/4$ is the dielectric permittivity of the cold plasma. The solution of Eq. (\ref{r}) can be written in the form
\begin{multline}
	\mathcal{E}=\left(\mathcal{A}-\frac{2i}{\sqrt{3}} \int\limits_{-l}^{y} \mathcal{J}(s) \cos(\sqrt{3}s) ds\right)\sin(\sqrt{3}y) \\
	+ \left(\mathcal{B}+\frac{2i}{\sqrt{3}} \int\limits_{-l}^{y} \mathcal{J}(s) \sin(\sqrt{3}s) ds\right)\cos(\sqrt{3}y).
\end{multline}
The constants $\mathcal{A}$ and $\mathcal{B}$ can be found from the boundary conditions at $y=\pm l$ by matching the internal plasma fields $E_x, B_z$ with the fields of radiated EM waves in vacuum
\begin{align}
	E_x&=\mathcal{C} e^{i 2y-i 2 \omega_p t}+c.c., \qquad y>l, \\
	E_x&=\mathcal{D} e^{-i 2y-i 2 \omega_p t}+c.c., \quad y<-l.
\end{align}
Thus, we can present the amplitude of radiated wave in the form
\begin{equation}\label{amp}
	\mathcal{E}_0=2\left|\mathcal{D}\right|=\frac{2}{\sqrt{\cos^2(\sqrt{3}l)+3}} \left|\int\limits_{-l}^{l} \mathcal{J}(y)\cos(\sqrt{3}y) dy\right|.
\end{equation}
 Taking into account a slow dependence of wakes amplitudes on the longitudinal coordinate and fixing the finite size $\sigma_z$ of the plasma channel along the $z$-direction, we obtain the total radiation power:
\begin{equation}
	\frac{P}{P_0}=  \sigma_z \int\limits_{-\infty}^{\infty} \mathcal{E}_0^2 dx,
\end{equation}
where $ P_0=m_e^2 c^5/(4 \pi e^2)\approx 0.69\ \mbox{GW}. $
 
Let us consider the case when counterpropagating plasma wakes are excited by short $y$-polarized laser pulses with the central frequency $\omega_0$ and the envelope:
\begin{equation}
	E_y= E_{0s} \sqrt{\frac{\sigma_{0s}}{\sigma_s(x)}} e^{-y^2/\sigma_s^2(x)} \sin^2\left(\frac{\pi (t\pm x)}{2\tau}\right). 
\end{equation}
Here, we take into account the effect of laser diffraction resulting in spreading of each driver as $\sigma_s(x)=\sigma_{0s} \sqrt{1+x^2/\mathcal{R}_s^2}$, where $\mathcal{R}_s=\omega_0 \sigma_{0s}^2/2$ is the Rayleigh length corresponding to the focal spot size $\sigma_{0s}$ ($t$ is measured in $\omega_p^{-1}$ units). In this plane geometry, we consider the case of large $\sigma_z$ ($\sigma_z\gg \sigma_s$) which is assumed constant along the filament. The amplitudes of excited wakefields should follow the same varying transverse structure that is determined by the laser-induced ponderomotive force
\begin{equation}
	E_s(y)=E_s^w \left(\frac{\sigma_{0s}}{\sigma_s(x)}\right) e^{-2y^2/\sigma_s^2(x)},
\end{equation}
where 
\begin{equation}
	E_s^w=\frac{3}{4} \frac{E_{0s}^2}{\omega_0^2} \frac{\sin \tau}{(4-5 \tau^2/\pi^2+\tau^4/\pi^4)}
\end{equation}
are the maximal longitudinal electric fields inside the waists of laser beams. This formula shows that the most efficient excitation of plasma wakefields is achieved for some optimal laser duration ($\tau\sim \pi$). Since the created plasma channel is wider than the wakefield size $\sigma_s$, the integration region in (\ref{amp}) can be considered as  infinite one. In this case, the integral (\ref{amp}) can be calculated analytically and the amplitude of radiated EM wave takes the form
\begin{align}\label{e0}
	&\mathcal{E}_0=\frac{3}{2}\sqrt{\frac{\pi}{2}} \frac{E_1^w E_2^w \mathcal{F}_{\sigma}}{\sqrt{\cos^2(\sqrt{3}l)+3}},  \\ 
	\mathcal{F}_{\sigma}=&\frac{\sigma_{01} \sigma_{02} \left|\sigma_2^2-\sigma_1^2\right|}{(\sigma_1^2+\sigma_2^2)^{3/2}} \exp\left[-\frac{3}{8} \frac{\sigma_1^2 \sigma_2^2}{\sigma_1^2+\sigma_2^2}\right].
\end{align}

There are several different ways how to implement this generating scheme in laboratory experiments. It is possible to focus counterpropagating laser pulses in a homogeneous gas either to the  spots with different sizes in a single $x$-point (Fig. \ref{f1}b) or to the equal spots spaced by the distance $L$ in the longitudinal direction (Fig. \ref{f1}c). In the former case, the region of intense radiation is localized near the single focus, whereas in the latter case each focus can radiate its own terahertz pulse.
For the plasma density $n_0=3\cdot 10^{18} \ \mbox{cm}^{-3}$ (radiation frequency $\nu\approx 31$ THz), a laser pulse with the wavelength 810 nm, total energy 1.6 J,  optimal duration $\tau\approx 31$ fs and maximal strength parameter $a_1=E_{01}/\omega_0=0.7$ can drive the linear plasma wave with the peak amplitude $E_1^w\approx 0.2$ inside a rather large plasma volume restricted by the optimal spot size $\sigma_{01}=1.38\ c/\omega_p$, $\sigma_z=417 c/\omega_p$ (determined by the total laser pulse energy) and longitudinal size $6 \mathcal{R}_1\approx 136\ c/\omega_p$. 
In the first scheme, the highest radiation fields with the amplitude $\mathcal{E}_0\approx 13$ MV/cm are achieved, when the counterpropagating laser with the same energy and same size $\sigma_z$ is focused to the wider spot in $y$-direction $\sigma_{02}=2.3 \sigma_{01}$. In this case, the total radiation power reaches 0.56 GW and, by the moment $100\ \omega_p^{-1}\approx 1$ ps, about 0.57 mJ of energy is radiated with terahertz waves, which corresponds to the energy conversion efficiency $\eta=1.8\cdot 10^{-4}$. In the second scheme, for the optimal distance between the focal spots $L=2.15 \mathcal{R}_1$, the same field $\mathcal{E}_0\approx 13$ MV/cm is generated in almost twice larger area, which results in producing two pulses with the total power 1 GW,  total energy 1 mJ and the efficiency $3.2\cdot 10^{-4}$.

\begin{figure}[htb]
\bc\includegraphics[width=240bp]{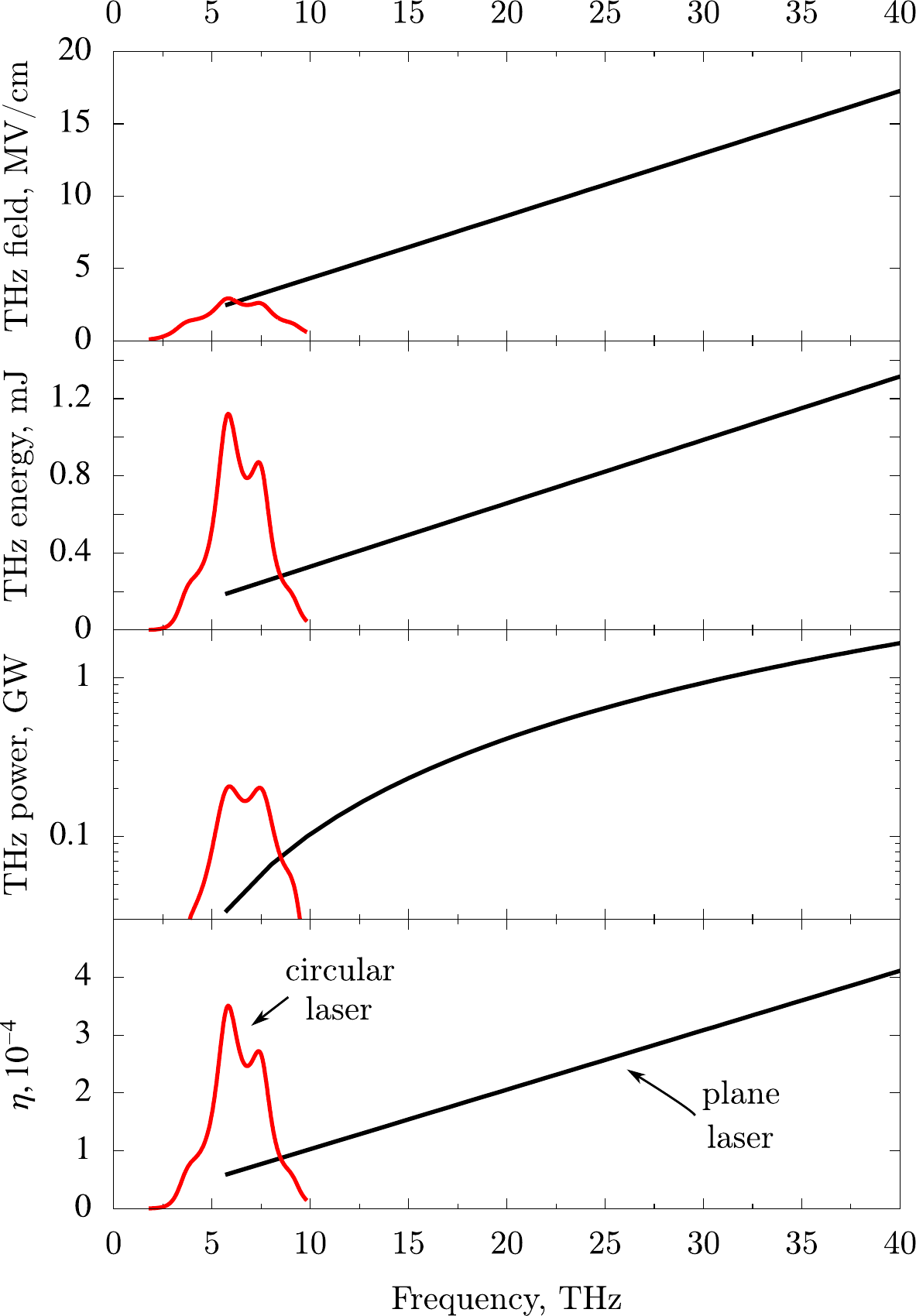} \ec \caption{Efficiency, energy, power and maximal field of THz radiation in scheme 2 as functions of radiation frequency.}\label{f2}
\end{figure}

The frequency of emitted radiation can be easily tuned by the change of plasma density $n_0$. To obtain the most efficient laser-to-THz conversion in the second generating scheme, varying the density, one should also vary the laser duration ($\sim 3/\omega_p$), focal spot sizes ($\sigma_0\approx 1.38\ c/\omega_p$ and $\sigma_z\propto \tau^{-1} \sigma_{0}^{-1}\propto n_0$) and distance $L\approx 2.1 \mathcal{R}\propto n_0^{-1}$ between laser beams waists. Varying all these quantities synchronously for the fixed values of laser pulse energy 1.6 J and maximal strength parameter $a_1=0.7$ and estimating the duration of THz emission at the level  $100\ \omega_p^{-1}$, we can predict how the laser-to-THz efficiency, THz energy, THz power and maximal THz field depend on the radiation frequency $\nu=\omega_p/\pi$ (black curves in Fig. \ref{f2}). 
In the low frequency region $\nu<5$ THz, $\sigma_z$ becomes comparable with the size $\sigma_{01}$ and the laser filament cannot be considered as a plane one  for the fixed laser energy. Figure \ref{f2} shows that the proposed generating scheme is able to produce high-field ($>10$ MV/cm) high-power ($\sim 1$ GW) terahertz pulses with the conversion efficiency $>2\cdot 10^{-4}$ in the frequency range $\nu>20$ THz, and less intensive pulses ($\sim 1$ MV/cm, $<100$ MW, $\eta< 1\cdot 10^{-4}$) at low frequencies 5-10 THz. This rapid decrease of radiation efficiency in rarefied plasmas is explained by less efficient excitation of plasma wakefields by laser drivers. 
\begin{figure*}[htb]
\bc\includegraphics[width=460bp]{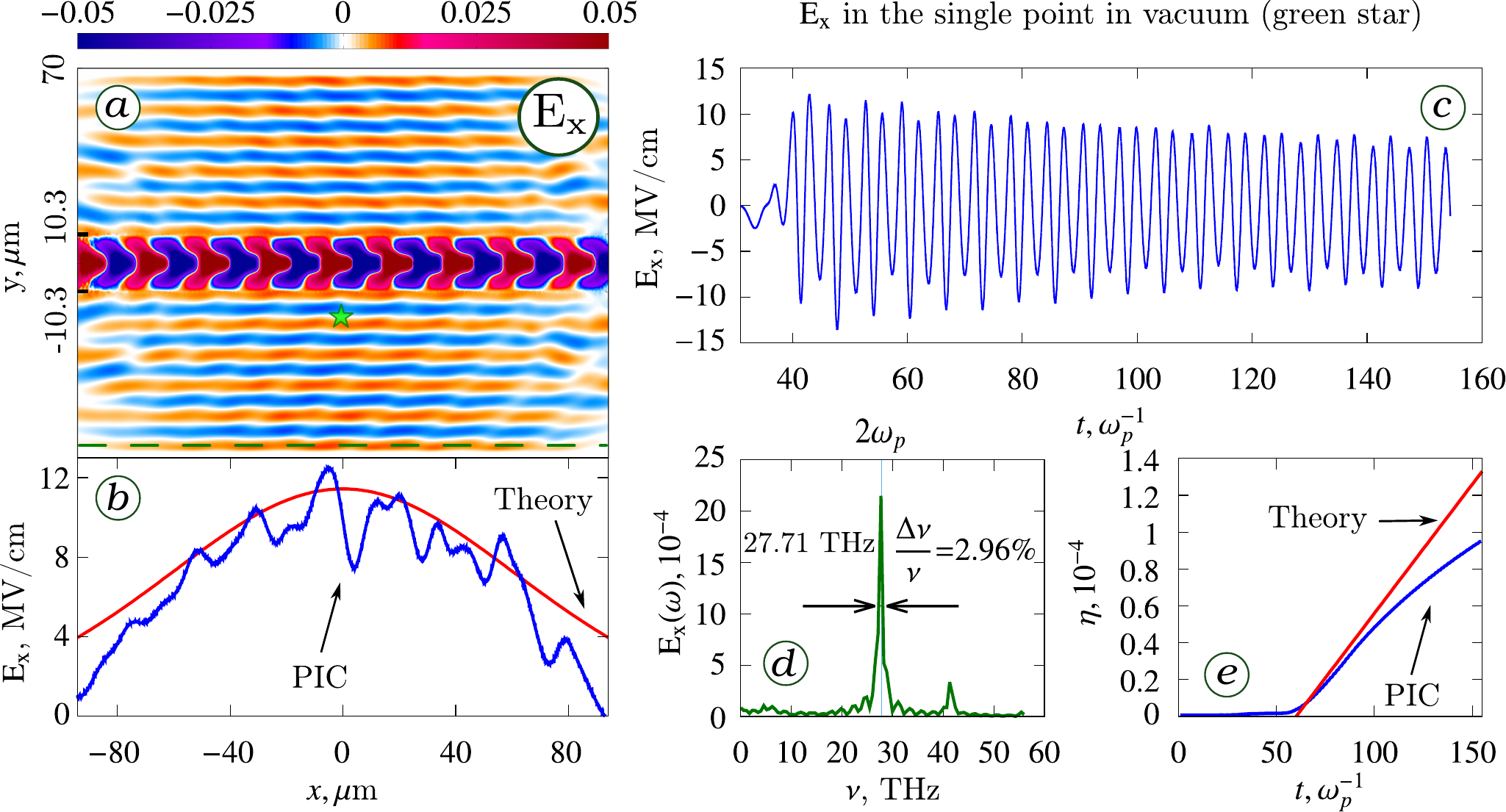} \ec \caption{Results of PIC simulations for $n_0=2.4\cdot 10^{18} \ \mbox{cm}^{-3}$. The map of dimensionless longitudinal electric field $E_x(x,y)$ in the moment $t=76.8 \omega_p^{-1}$ (a). The longitudinal profile of $E_x$ along the green dashed line (in MV/cm) and theoretical prediction for the amplitude $\mathcal{E}_0$ (\ref{e0}) (b). The history of THz field $E_x(t)$ in the single point indicated by the green star (c). The frequency spectrum of THz radiation (d). The laser-to-THz energy conversion efficiency as a function of time (e).}\label{f3}
\end{figure*}
It should be also noted that THz radiation lasts much longer than the chosen characteristic time $100\ \omega_p^{-1}$, that is why, in reality, the total energy conversion efficiency should be higher than that predicted in Fig. \ref{f2}. The reason why we consider such a short radiation time is the dissipation of energy concentrated in the main harmonics of plasma wakes due to excitation of non-radiating satellites. Our PIC simulations show that  $100\ \omega_p^{-1}$ is a typical time-scale at which the radiation power does not change drastically.

Since we do not have enough energy to create a plane laser filament at low frequencies $\leq 5$ THz, let us find out how radiation characteristics vary with the plasma density for the circular laser beam ($E_y\propto \exp(-r^2/\sigma_s^2)$). If the peak laser strength parameter in the waist is fixed at the constant level $a_1=0.7$, the focal spot radius $\sigma_{01}$ is completely determined by the laser energy. In cylindrical geometry, the amplitude of THz electric field achieved at the plasma boundary ($r=R$) can be written in the form
\begin{align}
	&\mathcal{E}_0= \frac{ 6 E_1^w E_2^w \mathcal{F}_{\sigma}}{\sqrt{(J_0+2\sqrt{3} R J_1)^2+16 R^2 J_0^2}},  \\ 
	\mathcal{F}_{\sigma}=&\frac{\sigma_{01}^2 \sigma_{02}^2 \left|\sigma_2^2-\sigma_1^2\right|}{(\sigma_1^2+\sigma_2^2)^{2}} \exp\left[-\frac{3}{8} \frac{\sigma_1^2 \sigma_2^2}{\sigma_1^2+\sigma_2^2}\right],
\end{align}
where $J_n=J_n(\sqrt{3}R)$ are the Bessel functions. The total radiation power is then determined by the integral $P/P_0=\pi R \int \mathcal{E}_0^2 dx$ over the longitudinal coordinate $x$. Red curves in Fig. \ref{f2} show that, for the optimal distance between the waists $L\approx 1.3 \mathcal{R}_1$, optimal duration $\tau\approx 3/\omega_p$ and for the plasma radius $R=2\sigma_{01}$, efficient generation of THz radiation in scheme 2 occurs in a rather narrow frequency region $4-7$ THz. It is explained by the existence of the optimal spot size which is tied to $c/\omega_p$. Thus, 1.6 J cylindrical laser beams focused to the spots with the radius 28 $\mu$m at the plasma density $n_0=10^{17}\ \mbox{cm}^{-3}$ can produce 1 mJ, 200 MW terahertz pulses with the efficiency $3.5\cdot 10^{-4}$. It is interesting to note that generation of THz radiation can be more efficient for less energetic lasers. Our theory predicts that, if laser beams with the same energy 15 mJ and duration 35 fs are focused to the circular spots with the diameter 12 $\mu$m spaced by 180 $\mu$m, radiation efficiency for 27 THz reaches $0.16\%$. 

In order to confirm that the proposed mechanism of EM waves generation can provide high-field narrowband terahertz pulses, we perform 2D3V particle-in-cell simulations. Our simulation box consists of a narrow plasma slab ($2l=6\ c/\omega_p$) surrounded by vacuum gaps and boundary layers absorbing the radiated EM waves. In our model, plasma wakefields are excited by the virtual counterpropagating laser pulses acting on plasma electrons through the ponderomotive force only. Such a simplified approach allows us to use the relatively large spatial and temporal grid steps ($\Delta x=0.01\ c/\omega_p$ and $\Delta t=0.005\ \omega_p^{-1}$) and relatively small number of macroparticles ($N_p=3.2\cdot 10^{7}$).  To compare with theoretical predictions, we simulate the scheme 1 in which laser drivers with peak strengths $a_1=0.7$ and $a_2=a_1 \sqrt{\sigma_{01}/\sigma_{02}}$ are focused to the spots $\sigma_{01}=1.38\ c/\omega_p$ and $\sigma_{02}=2.3 \sigma_{01}$ situated in the center of the simulation box. Fig. \ref{f3}a shows that the nonlinear interaction of laser-driven plasma waves ($E_1^w=0.19$ and $E_2^w=0.08$) does really generate the second harmonic EM emission in which the amplitude of electric field $E_x$ (measured along the damping layer) agrees well with the theoretical profile (\ref{e0}) (Fig. \ref{f3}b). Some deviation from this profile near the boundaries is explained by the fact that lasers widths in these regions slightly exceed the size of the preformed plasma channel.  From Fig. \ref{f3}c, one can see the history of $E_x$-field measured in the single point in vacuum indicated by the green star. It confirms the multi-cycle nature of radiation which is concentrated inside a narrow spectral line with the relative width $\Delta \nu/\nu<3\%$ (Fig. \ref{f3}d). Fig. \ref{f3}c also demonstrates a slow decrease of the THz amplitude in time, which becomes more visible in the temporal dependence of the total radiated energy shown in Fig. \ref{f3}e.
This decrease, compared to the theoretical linear growth, is caused by the dissipation of energy stored in the dominant wake harmonics. Besides the radiation losses ($7\%$), these harmonics also transfer the energy to a number of non-radiating satellites $(2\omega_p,\pm 2k)$ and $(0,\pm 2k)$. In the absence of counterpropagating waves, the plasma wakes do not lose their energy during the simulation time.

In conclusion, we propose a new scheme for the generation of high-field ($1-20$ MV/cm) tunable ($4-40$ THz) narrowband ($<3\%$) THz radiation by counterpropagating plasma wakefields. Using a simple theory, we calculate the power, total energy and conversion efficiency of the THz radiation generated by plane and circular joule-scale femtosecond lasers  and confirm these predictions by PIC simulations. It is shown that this generating scheme can provide GW, mJ multi-cycle terahertz pulses with the efficiency $>10^{-4}$ ($>10^{-3}$ for low-energy lasers), which opens novel opportunities in selective nonlinear control of matter and other applications.  

Authors thank Prof. K.V.Lotov and Prof. A.P.Shkurinov for fruitful discussions.
This work is financially supported by the Russian Foundation of Basic Research (grant 15-32-20432). Simulations are performed using the resources of Novosibirsk State University.

\end{document}